\documentclass[a4paper]{jpconf}
\usepackage{graphicx}

\begin{document}
\title{The magnetic environment in the central region of nearby galaxies}

\author{Cornelia C Lang and Maria R Drout}

\address{Department of Physics \& Astronomy, University of Iowa, Iowa City, IA 52245, USA}

\ead{cornelia-lang@uiowa.edu}

\begin{abstract}
The central regions of galaxies harbor some of the most extreme physical phenomena,
including dense stellar clusters, non-circular motions of molecular clouds and strong
and pervasive magnetic field structures.  In particular, radio observations have shown that 
the central few hundred parsecs of our Galaxy has a striking magnetic field configuration. It
is not yet clear whether these magnetic structures are unique to our Milky Way or a common
feature of all similar galaxies.  Therefore, we report on (a) a new radio polarimetric survey 
of the central 200 pc of the Galaxy to better characterize the magnetic field structure and (b) a search 
for large-scale and organized magnetized structure in the nuclear regions of nearby galaxies using
data from the Very Large Array (VLA) archive. The high angular resolution (1--5$^{\prime\prime}$) 
of the VLA allows us to study the central 1\,kpc of
the nearest galaxies to search for magnetized nuclear features similar to what is
detected in our own Galactic center. Such magnetic features play a 
important role in the nuclear regions of galaxies in terms of gas transport 
and the physical conditions of the interstellar medium in this unusual region of galaxies. 
\end{abstract}

\section{Introduction}
\vskip 0.1in
Magnetic fields play crucial roles in the interstellar environments of galaxies and are observed on 
on a number of scales in galaxies, up to scale lengths of 8\,kpc for the regular, organized 
fields in the disks (e.g., Beck 2004, Beck and Gaensler 2006). Radio synchrotron observations
are one of the best ways to probe magnetic fields across galaxies as radio waves are not subject to
interstellar extinction. In addition to revealing maps of the magnetic field structure, radio observations
give estimates for the mean equipartition magnetic field strengths in galaxies from $\sim$ 10-20 $\mu$G in 
galactic spiral arms, and $\sim$40 $\mu$G in galactic nuclear regions. 
The central regions of galaxies harbor some of the most extreme physical phenomena,
including dense stellar clusters, non-circular motions of molecular clouds and strong
and pervasive magnetic field structures. In addition, studies across the electromagnetic spectrum reveal 
that galactic nuclei are major sites of episodic and energetic phenomena related both to active 
galactic nuclei (AGN) and also to intense bursts of star formation. 

Recently, studies of the circumnuclear regions of several galaxies have revealed that magnetic fields in this
part of the galaxy may help to transport materials into the nuclear region (which then may fuel a starburst or
AGN). In the barred galaxies, NGC\,1097 and NGC\,1365, radio emission is enhanced along
the bar regions, indicating the presence of a shock front. Further, magnetic stresses in the circumnuclear ring
can then drive mass inwards at rates high enough to feed nuclear activity (Beck \textit{et al} 1999; 2005).  
In the spiral ringed galaxy M\,94 (NGC\,4736) the polarized radio emission reveals a pattern of ordered magnetic
field arising from the central regions of the galaxy that may be where magnetic amplification occurs (Chyzy and Buta 2008).  
In addition, the expulsion of materials from the nucleus may also be regulated by magnetic fields. Energetic events 
arising from star-burst or AGN activity in the nucleus often result in explosive vertical ``fountains'' of 
million degree gas, rising 1--2\,kiloparsec (kpc) above galactic 
nuclear regions (e.g., NGC\,1569 and NGC\,4631; Heckman \textit{et al} 1995). These fountains may well be related to vertical
magnetic field structures and outflow from the galactic disk (e.g., NGC\,4631; Golla and Hummel 1994).  

Detailed knowledge of the violent interplay between stellar and magnetized interstellar 
components is limited for most galactic nuclei because of their great distances, even for the 
most sophisticated observational instrumentation.
At a distance of only 8.0\,kpc, the center of our Milky Way galaxy provides an excellent 
laboratory for detailed studies of the interplay of massive stars, gas and, importantly, magnetic fields. 
Because of the nearly 30 magnitudes of visual extinction toward this region of our Galaxy, 
studies at radio wavelengths have proved to be fruitful in understanding the details of 
the magnetized interstellar medium and for understanding the nuclei of nearby, normal galaxies.

\section{Our Galactic Center}
\vskip 0.1in
\subsection{The Magnetized Environment}
\begin{figure}[t!]
\includegraphics[clip,width=0.68\textwidth]{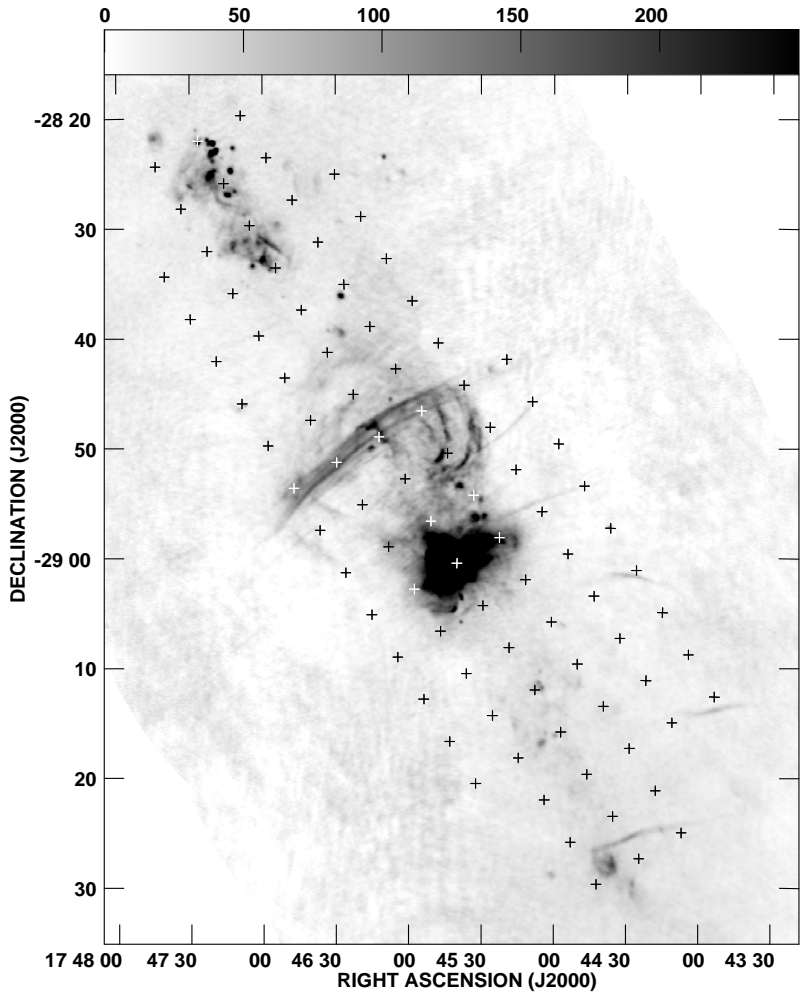}\hfill%
\begin{minipage}[b]{11pc}
\caption{\label{fig1}Very Large Array (VLA) 1.4\,GHz mosaic of the central 200 pc of the GC from 
the recent HI absorption study (Lang \textit{et al} {\it in prep.}). Crosses represent the positions of 90 pointings
made with the VLA at 4.9\,GHz in order to cover this same region with full-polarization imaging at $\sim9^{\prime\prime}\times5^{\prime\prime}$
angular resolution.}
\end{minipage}
\end{figure}

\vskip 0.1in
One of the most striking aspects of the radio continuum image of our Galactic center (GC) shown in 
Fig.\ \ref{fig1} is the presence of the narrow linear features oriented essentially perpendicular 
to the orientation of the Galactic plane. 
First detected with the VLA by Yusef-Zadeh, Morris and Chance (1984), the 8 well-known non-thermal filaments (NTFs) 
have the following properties 
(Gray \textit{et al} 1995; Yusef-Zadeh \textit{et al} 1997; Lang \textit{et al} 1999a): (1) show strong linear polarization (30$-$50\% in most cases) and
(2) have intrinsic magnetic field orientations aligned along their lengths,indicating that they may 
trace a large-scale poloidal magnetic field (e.g., Morris and Serabyn 1996). 
The magnetic field traced by these NTFs is opposite to the orientation 
of the magnetic field in the galactic disk, which is azimuthal and follows the spiral arms. Several 
theories have explored the origin and stability of such a field structure (Chandran, Cowley and Morris 2000; Chandran 2001). 
One idea is that the magnetic field may be pervasive throughout the GC, with the NTFs representing 
sites of relativistic particles (Morris 1994). More recently, many new NTFs have been detected with 
lower surface brightnesses, shorter extents and many different orientations (Lang \textit{et al} 1999b; 
LaRosa \textit{et al} 2004; Yusef-Zadeh \textit{et al} 2004) and the overall structure of the magnetic field
has yet to be fully uncovered. La Rosa \textit{et al} (2005) suggest that the numerous weak and randomly-oriented
filamentary structures suggest a much weaker, more local and dynamic field configuration in this
region of the Galaxy.

The magnetic pressure for field 
strengths of 0.1--1\,mG (which is what is estimated for the magnetic 
field in the GC) corresponds
to 4 $\times$10$^{-10}$ to 4 $\times$10$^{-9}$ erg cm$^{-3}$ and is likely to be balanced by the
pressure from the hottest interstellar components. 
Careful comparisons between diffuse X-ray emission and radio features can provide insight 
on this balance. The results
of Wang, Gotthelf and Lang (2002) show that the diffuse X-ray emission is consistent with 
T$\sim$10$^7$ K hot gas, which corresponds to a pressure in the hot component of 
$\sim$1$-$5$\times$10$^{-10}$ erg cm$^{-3}$ for a particle density of 0.1$-$0.5 cm$^{-3}$. Therefore, 
the correlation between the diffuse, hot X-ray emission and the magnetic features in the GC
region has important implications for the confinement of hot gas, and the ultimate transport of
such energetic ISM components out of the nuclear region of the Galaxy (e.g., Shibata and Uchida 1987).

\subsection{Large-scale VLA 4.9\,GHz Polarimetric Survey}
\vskip 0.1in
Although a number of lower frequency surveys 
of the GC have recently been made (Nord \textit{et al} 2004; Yusef-Zadeh \textit{et al} 2004), uniform coverage
at a higher frequency ($>$ 1.4\,GHz) has not been carried out until this work. In particular, large rotation 
measures ($> 1000$~rad~m$^{-2}$) toward the GC (Yusef-Zadeh \textit{et al} 1997; Lang \textit{et al} 1999a,b) cause complete
depolarization at~1.4\,GHz, which is why this 4.9\,GHz survey is ideally
suited for detecting polarized intensity from magnetized features in the GC, including the
enigmatic new ``streaks'' (NTF-candidates) (Nord \textit{et al} 2004; Yusef-Zadeh \textit{et al} 2004). 
Determining whether the new NTF-candidate sources are polarized may allow us to uncover an underlying 
magnetic field structure in the GC and will greatly increase the number of NTF sources in this region. 

We have begun to construct the first 4.9\,GHz mosaic of the GC region made from 90 pointings in 
D-configuration and 50 pointings in C-configuration with the VLA in order to study extended 
structures in total and polarized intensity. Observations were made during 2003-2006, and data calibration for
all fields in both configurations is nearly completed. The quality of the data is very high and the total integration
time on source is $\sim$75 minutes per pointing. The combination of data from these two configurations should 
provide sensitivity to diffuse large-scale structures but also have a final angular resolution of $<$10$^{\prime\prime}$.
Mosaicking such a large and complex region in full Stokes mode (i.e., both total and polarized intensity) 
has proven to be a challenge.  The Multi-Scale Clean algorithm available
in {\it CASA}, the new software package being developed by the National Radio Astronomy Observatory
is designed for fields like our GC mosaic, which have a substantial diffuse component 
in addition to numerous compact sources. 

\begin{figure}[t]
\begin{center}
\includegraphics[clip,height=0.8\textheight]{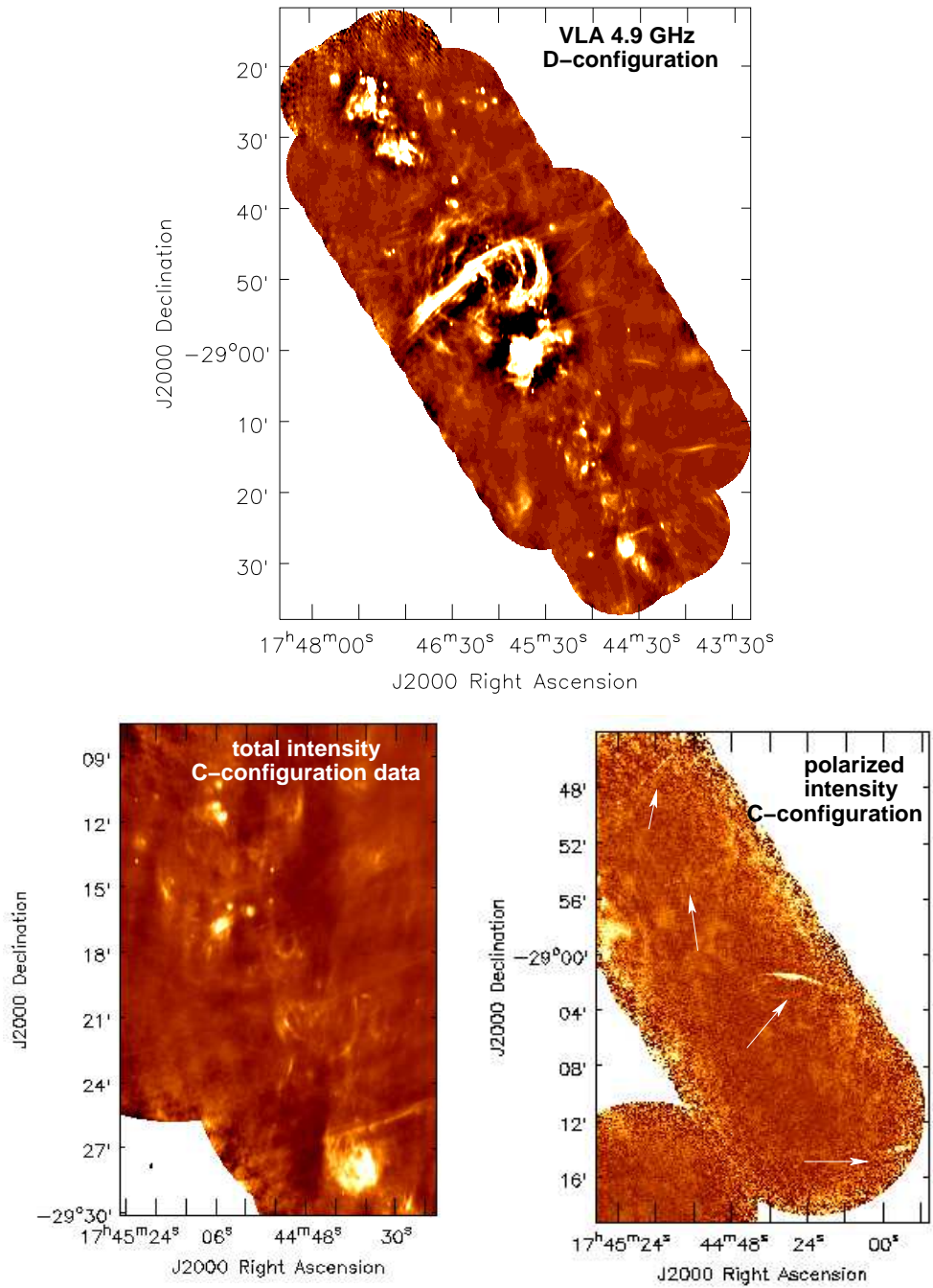}
\end{center}
\caption{\label{fig2}(Top) Preliminary 4.9\,GHz mosaic image using D-array data only (resolution $\sim$20$^{\prime\prime}$). (Lower left): Inset of
C-array mosaic in the vicinity of SgrC showing numerous compact and shell-like features indicating the presence of massive star forming 
activities with resolution $\sim6^{\prime\prime}$. (Lower right): Inset of polarized intensity from the C-array mosaic, with 4 NTFs at positive 
Galactic latitudes labeled with white arrows.}
\end{figure}

Figure \ref{fig2} shows preliminary images from both the D- and C-configuration data
in total and polarized intensity. The preliminary results are promising, revealing newfound detail
in several areas: (1) numerous compact and extended features in the region between 
Sgr\,A and Sgr\,C, shown in Fig.\ \ref{fig2} (lower left), have not been previously studied in
the radio continuum with such high resolution. A number of these sources appear 
to have shell-like morphology which indicates they may be tracing massive star-forming activities. 
Several of these sources have mid-IR counterparts in recently published {\it Spitzer} images (Stolovy \textit{et al} {\it in prep.}); 
(2) the survey has begun to reveal polarized intensity from several candidate NTFs, and upon further polarization corrections, we hope to uncover
many more polarized counterparts to candidate NTFs. Figure \ref{fig2} (lower right) shows strong polarized intensity from four of 
the well-known NTFs; finally,  (3) the C-configuration data show a large number of new compact sources which will assist in
understanding the true radio point source population in the GC and the nature of these sources (e.g., pulsars, extragalactic, etc.).   
When completed, by combining the D- and C-configuration data and producing a complete source catalog for total and polarized intensity, 
this survey should be able to address the following questions: (1) Are the numerous NTF-candidates polarized, and if so, do they trace out
an underlying complex magnetic field structure, (2) Is the magnetic field in the GC region strong and
pervasive or weak and diffuse with enhancements? (3) How do massive stars impact the ISM? and (4) What is the radio
point-source population at the GC? 

\section{Nearby Galactic Nuclei}
\vskip 0.1in
Our results from the GC indicate that the magnetic field
is strong and well-organized in this region and plays an important role in the transport of plasma
and energy out of the nuclear region. We have therefore begun an archival investigation of the central regions of the nearest, 
normal-type galaxies (analogs to the Milky Way). A search for similarly coherent magnetized features in 
nearby galaxies will allow us to better understand the role and uniqueness of the magnetic fields in the GC of
the Milky Way. 

\subsection{Target Criteria}
\vskip 0.1in
The best targets for such a study are located at distances $<$ 10\,Mpc, so that 1$^{\prime\prime}$~(which is 
a typical resolution of the VLA) resolution corresponds to no more than about 50 pc. In addition,
potential targets should be normal galaxies not thought to be undergoing a large starburst episode 
(although may have in the past). The study of extended, polarized features in M\,81 illustrates that
the nuclear regions of other normal galaxies may be rich in magnetized and filamentary features similar
to those found in our own Galactic center(Kaufman \textit{et al} 1996). These authors used the VLA in all its
configurations at both 1.4 and 4.9\,GHz to obtain 1$^{\prime\prime}$~and better spatial resolution. Kaufman \textit{et al} (1996) 
identify a large-scale (up to 1\,kpc) magnetized ``arc" within the inner 1\,kpc of the galaxy, indicating that
the interstellar magnetic field in this region is well-organized. In addition, the compact source in the center of 
M\,81 (M\,81$^*$; Brunthaler \textit{et al} 2001) has been found to be one of the best analogs to the GC compact source Sgr\,A$^*$. 
Sjouwerman \textit{et al} (2005) have used a large number of 4.9\,GHz multi-configuration, archival, VLA observations of the
nuclear regions of M\,31 and identified a non-thermal filamentary structure at very weak flux levels in the central 
kpc of the galaxy. Using the VLA archive, we identified multi-configuration, multi-frequency radio continuum 
observations of the nuclear regions of the above galaxies at 1.4 or 4.9\,GHz in the A, B and C array
configurations.   

\subsection{Archival Imaging \& Results}
\begin{figure}[t]
\begin{minipage}{22pc}
\includegraphics[width=22pc]{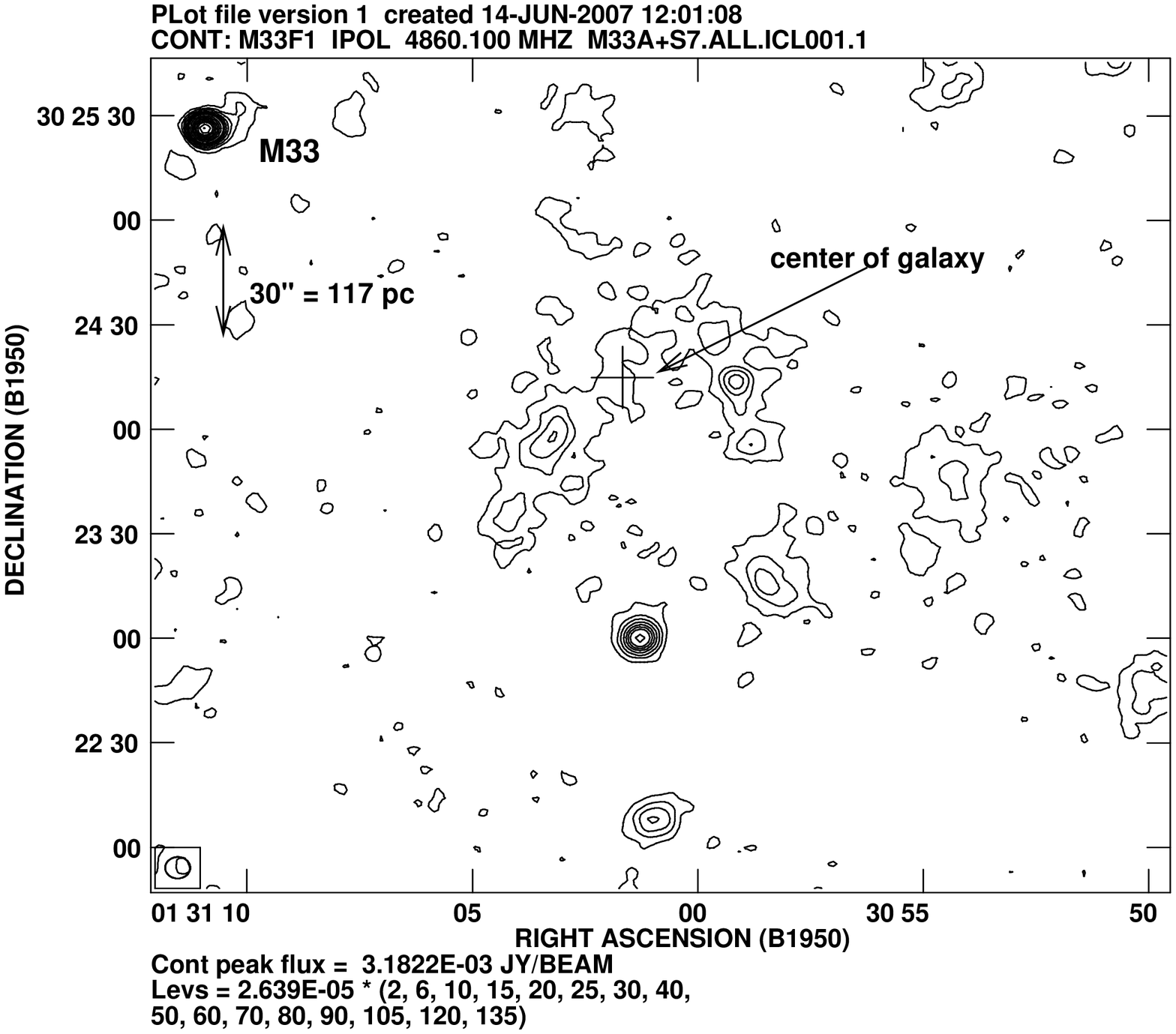}
\caption{\label{fig3} Nuclear region of M\,33 at 4.9\,GHz (resolution: 
$7.17^{\prime\prime}\times6.16^{\prime\prime}$).}
\end{minipage}\hfill
\begin{minipage}{14pc}
\includegraphics[width=14pc]{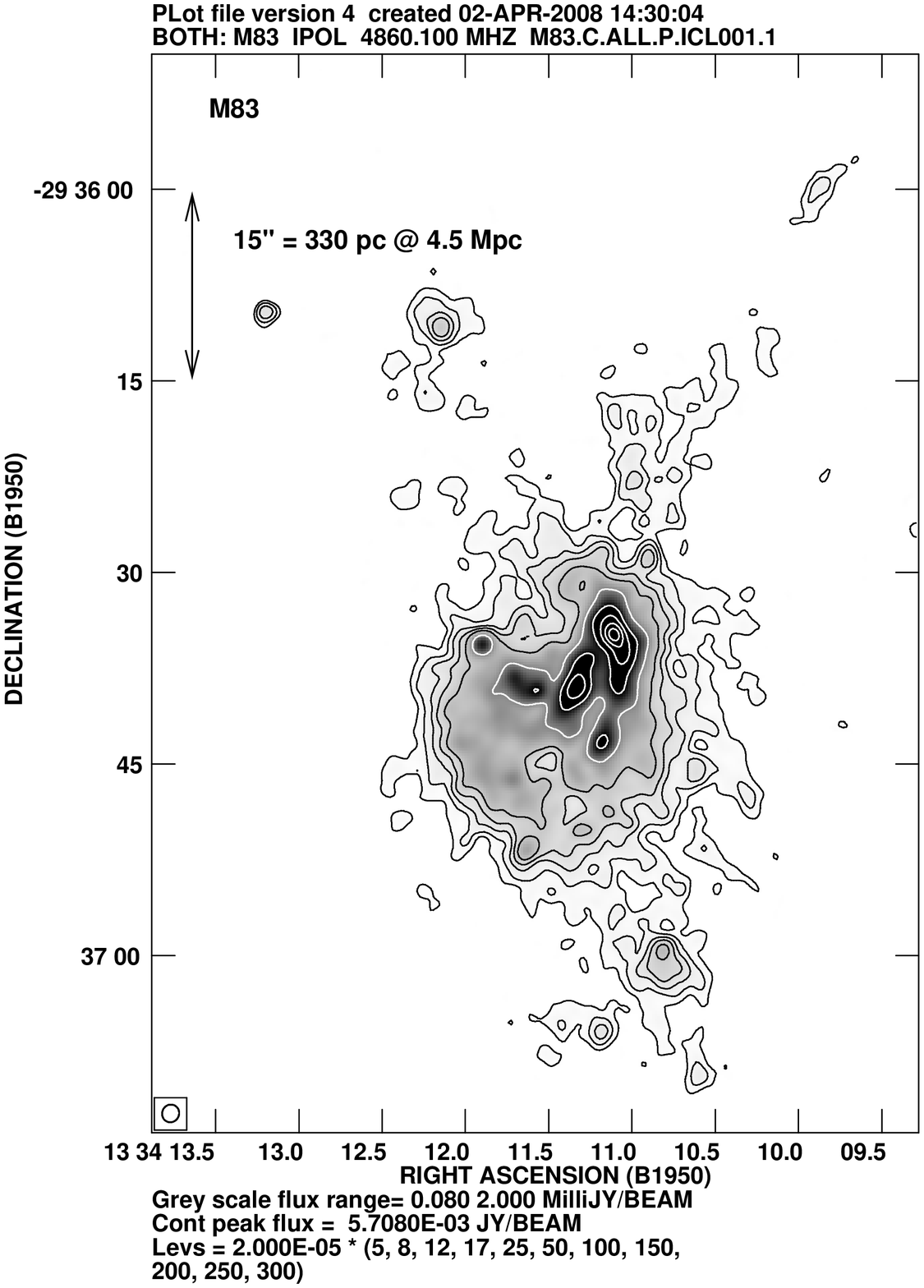}
\caption{\label{fig4}Nuclear region of M\,83 at 4.9\,GHz (resolution: 
$1.40^{\prime\prime}\times1.32^{\prime\prime}$).}
\end{minipage} 
\end{figure}

\begin{figure}[t]
\begin{minipage}{19pc}
\includegraphics[width=19pc]{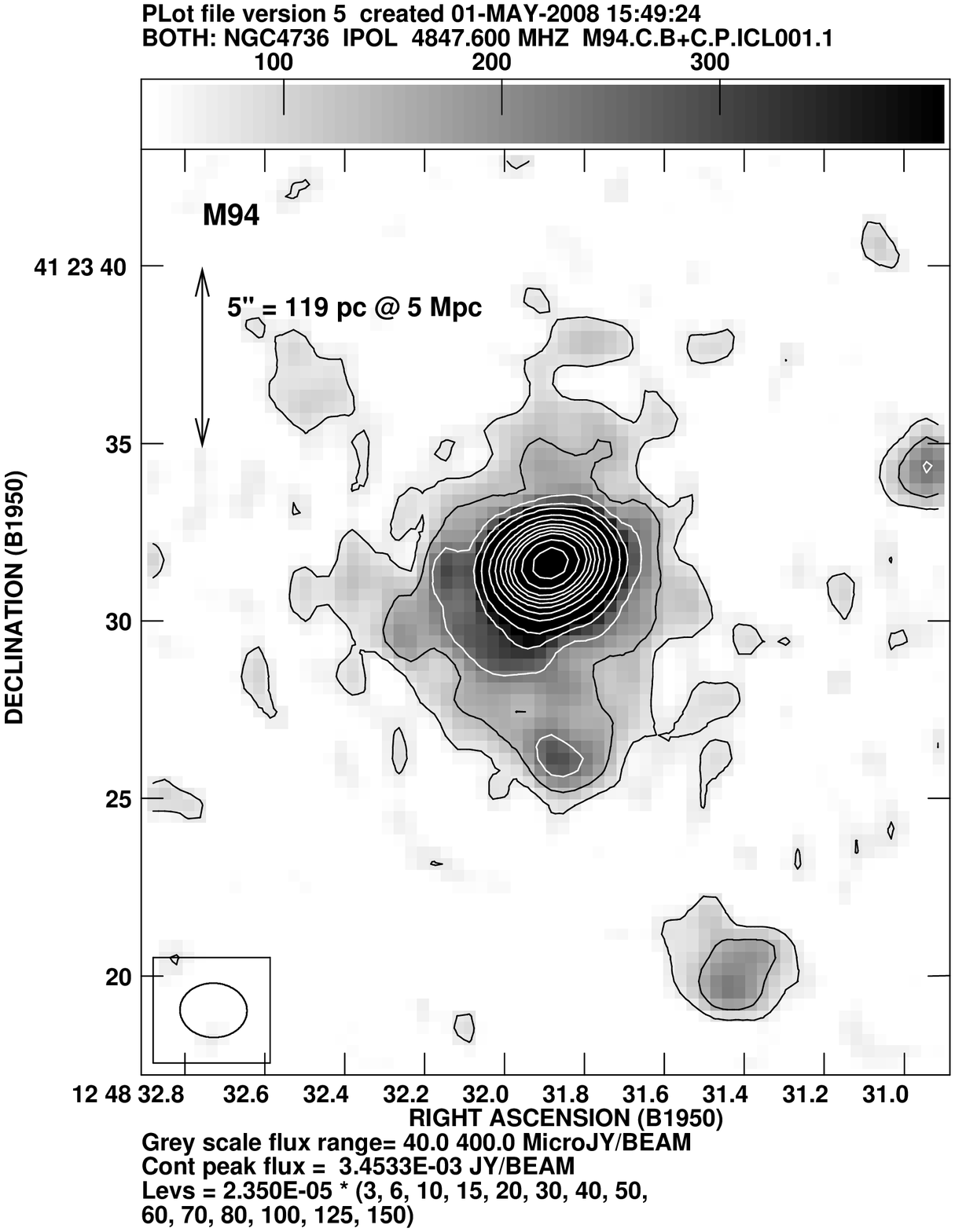}
\caption{\label{fig5}Nuclear region of M\,94 at 4.9\,GHz (resolution: 
$1.89^{\prime\prime}\times1.53^{\prime\prime}$).}
\end{minipage}\hfill
\begin{minipage}{18pc}
\includegraphics[width=18pc]{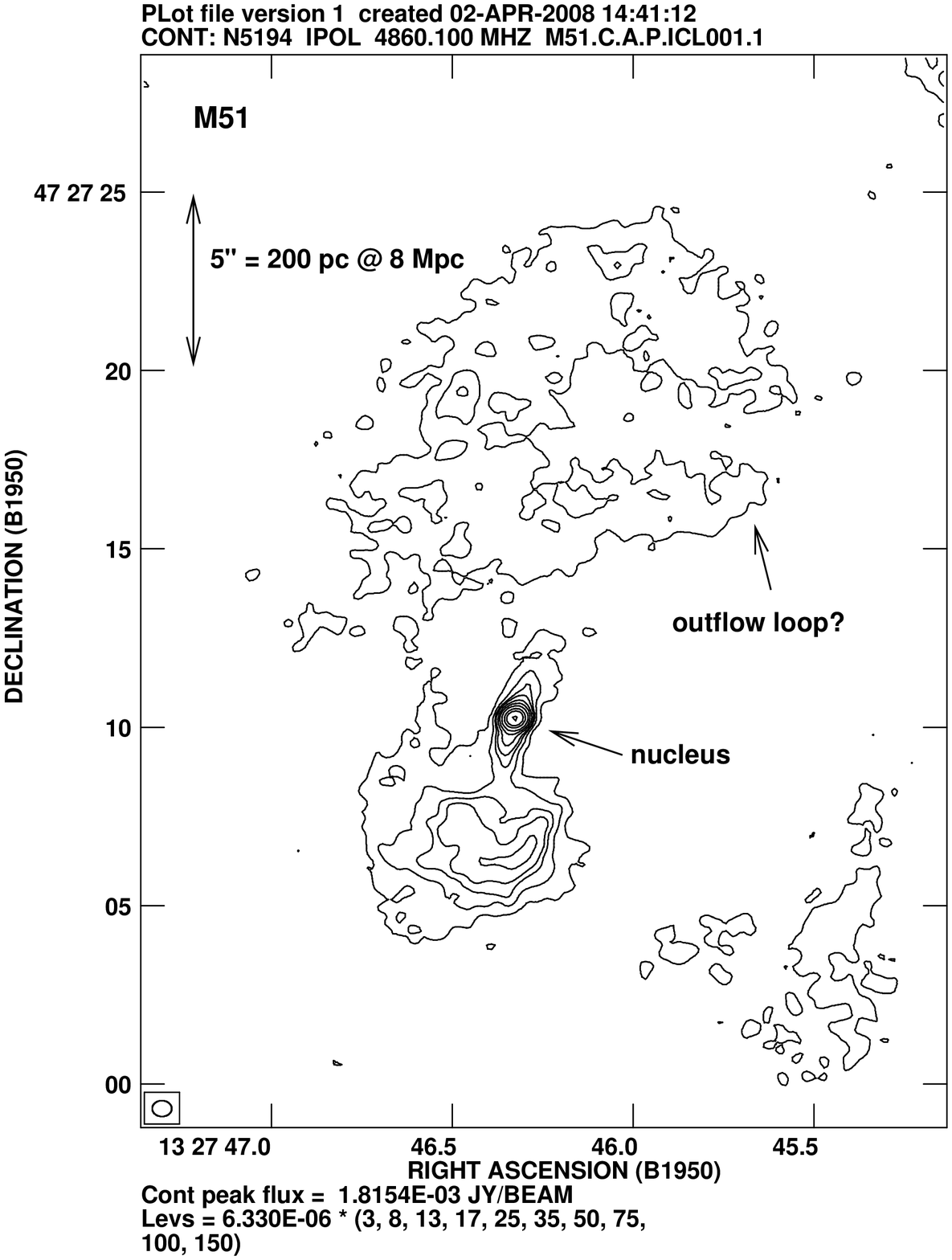}
\caption{\label{fig6}Nuclear region of M\,51 at 4.9\,GHz (resolution:  0.55$^{\prime\prime}$$\times$0.45$^{\prime\prime}$).}
\end{minipage} 
\end{figure}

\vskip 0.1in
Table 1 lists the galaxies for which we found data at 1.4 and 4.9\,GHz with sufficient angular
resolution ($<$5$^{\prime\prime}$) and high sensitivity (rms levels $<$ 0.1 mJy) in the VLA archive.  
Figures \ref{fig3}--\ref{fig6} show images of the central kpc or less in each of the galaxies for which we found 
archival observations. In all cases, there was sufficient {\it (u,v)} coverage and integration
time to do polarization calibration. 

\begin{table}[t]
\caption{\label{opt}Nearby Galactic Nuclei Properties}
\centering
\begin{tabular}{@{}*{7}{l}}
\br
Galaxy&Distance&1$^{\prime\prime}$~Scale&Reference/Archive Code\\
\mr
M\,31&d=0.7 Mpc&1$^{\prime\prime}$=3 pc&Sjouwermann \textit{et al} (2005)\\
M\,33&d=0.8 Mpc&1$^{\prime\prime}$=4 pc& this paper/AK140\\
M\,81&d=3.5 Mpc&1$^{\prime\prime}$=17 pc&Kaufman \textit{et al} (1996)\\
M\,83&d=4.5 Mpc&1$^{\prime\prime}$=22 pc& this paper/AW418\\
M\,94&d=4.5 Mpc&1$^{\prime\prime}$=22 pc&this paper/AD145\\
M\,51&d=8.0 Mpc&1$^{\prime\prime}$=40 pc&this paper/AC147\\
\br
\end{tabular}
\end{table}

\vskip 0.1in
\noindent {\bf M\,33:} The 4.9\,GHz image of M\,33 in Fig.\ \ref{fig3} shows the central few hundred parsecs of M\,33. There is
diffuse emission in an 'arc'-like configuration surrounding the nucleus. This emission extends for approximately
 1.5' total (350 pc at 0.8\,Mpc), or $\sim$150 pc on either side of the nucleus (indicated by a cross). Using a 
similar-resolution 1.4\,GHz image, the spectral index of the radio emission was found to vary between 
$\alpha$=$-$0.2 and $-$0.4, roughly consistent with non-thermal emission. Total intensity in this diffuse 
arc is faint (100--300\,$\mu$Jy\,beam$^{-1}$) and the rms level in the image is 
$\sim$25 $\mu$Jy beam$^{-1}$. Therefore, for even 50\% polarization ($\sim$50-150 $\mu$Jy beam$^{-1}$), which is
optimistic, the sensitivity in this dataset is not high enough to make a confident detection of polarized intensity. 
\vskip 0.1in
\noindent {\bf M\,83:} Figure \ref{fig4} shows radio emission at 4.9\,GHz arising from the central 1\,kpc of M\,83 with $\sim$1$^{\prime\prime}$ resolution. 
Diffuse radio emission is present over the entire nuclear region, with peaks near the very center of
the galaxy. The extended emission has a spectral index between 4.9 and 1.4\,GHz of $\sim$$-$0.5 to $-$0.9. There is
no detectable polarization arising from the brightest regions of total intensity ($\sim$2$-$4 mJy beam$^{-1}$), where it
is possible to detect down to $\sim$10\% polarization; however, for some of the weaker structures, it may not be possible
to detect polarization with these archive data. 

\vskip 0.1in
\noindent {\bf M\,94:} Figure \ref{fig5} shows the central 300\,pc of the nucleus of M\,94 with a resolution of $\sim$2$^{\prime\prime}$. 
The radio emission shows a non-thermal spectral index in this region with $\alpha$=$-$0.5. 
Weak polarization (signal-to-noise of $\sim$3) was detected at 4.9\,GHz in the central regions of the 
source, but was not detected at 1.4\,GHz, so we can not confirm that the archival data show polarization. 
However, recent VLA observations in D-configuration at 8 and 5\,GHz with 
10--15$^{\prime\prime}$ resolution show that the nuclear
region is strongly magnetized with a well-ordered circumnuclear field on slightly larger scales (Chyzy and Buta 2008). 

\vskip 0.1in
\noindent {\bf M\,51:} A 1$^{\prime\prime}$ image of the central regions of M\,51 is shown in Fig.\ \ref{fig6}. The nuclear source is the
bright point source in the center of the image. The nuclear source does not show polarization at 4.9\,GHz, with 
rms noise levels of 15\,$\mu$Jy. The source to the north of the nucleus has a loop-like morphology and is strongly 
suggestive of an outflow (Maddox \textit{et al} 2007). In Fig.\ \ref{fig6}, this source has very weak radio emission (20\,$\mu$Jy 
or so), and appears to be non-thermal in nature; however, these data do not have high enough sensitivity 
to reliably detect a polarized intensity counterpart for the loop-like structure. 

\section{Future Work \& Instrumentation}

The VLA archive search was a useful first attempt and allowed us to determine that 
existing observations are not sensitive enough to do a study of the polarization properties
of diffuse emission in the central hundreds of parsecs to 1\,kpc in nearby galaxies. 
Since sensitivity appears to be the main challenge, this project is ideally suited for the
Expanded VLA (EVLA). This upgrade to the VLA is currently in progress and expected to be
fully available by 2012; in the meantime, there are many opportunities for shared-risk observing
as the receivers are upgraded and correlator tested, and eventually replaced. The EVLA is
expected to have a factor of 10 improvement in sensitivity, especially in the 1-10\,GHz range. 
This will be crucial for a study like this one, where we are hoping to obtain high-sensitivity
images of nearby nuclei. 

\ack 
C.C.L. and M.R.D. acknowledge support from the Math and Physical Sciences Funding Program at University of Iowa for
this work. In addition, C.C.L. wishes to thank the International Travel Faculty Travel Award program at the 
University of Iowa for supporting her trip to the AHAR\,2008 conference to present these results.

\section*{References}

\medskip
\begin{thereferences}{}

\item Beck R, Ehle M,  Shoutenkov V, Shukurov A and Sokoloff D 1999 {\it Nature} \textbf{397} 324 

\item Beck R 2004 {\it Astrophysics and Space Sciences} \textbf{289} 293 

\item Beck R and Gaensler B 2004 {\it New Astronomy Reviews} \textbf{48} 1289

\item Beck R, Fletcher A, Shukurov A, Snodin A, Sokoloff D~D, Ehle M, Moss D, and Shoutenkov V 2005 {\it Astronomy and Astrophysics} \textbf{444} 73

\item Brunthaler A, Bower G~C, Falcke H and Mellon R~R 2001 {\it Astrophysical Journal} \textbf{560} L123 

\item Chandran B 2001 {\it Astrophysical Journal} \textbf{562} 737

\item Chandran B, Cowley S and Morris M 2000 {\it Astrophysical Journal} \textbf{528} 723

\item Chy{\.z}y K~T, and Buta R~J 2008 {\it Astrophysical Journal} \textbf{677} L17 

\item Golla G, and Hummel E 1994 {\it Astronomy \& Astrophysics} \textbf{284} 777 

\item Gray A~D, Nicholls J, Ekers R~D and Cram L~E 1995 {\it Astrophysical Journal} \textbf{448} 164 

\item Heckman T, Dahlem M, Lehnert M, Fabbiano G, Gilmore D and Waller W 1995 {\it Astrophysical Journal} \textbf{448} 98

\item Kaufman M, Bash F~N, Crane P~C and Jacoby G~H 1996 {\it Astronomical Journal} \textbf{112} 1021


\item Lang C~C, Morris M and Echevarria L 1999a {\it Astrophysical Journal} \textbf{525} 727

\item Lang C~C, Anantharamaiah K~R, Kassim N and Lazio T~J~W 1999b {\it Astrophysical Journal} \textbf{521} L41

\item LaRosa T~N, Nord M~E, Lazio T~J~W and Kassim N~E 2004 {\it Astrophysical Journal} \textbf{607} 302 

\item LaRosa T~N, Brogan C~L, Shore S~N, Lazio T~J, Kassim N~E and Nord M~E 2005 {\it Astrophysical Journal} \textbf{626} L23 

\item Maddox L~A, Cowan J~J, Kilgard R~E, Schinnerer E and Stockdale C~J 2007 {\it Astronomical Journal} \textbf{133} 2559

\item Morris M 1994 {\it The Nuclei of Normal Galaxies: Lessons from the Galactic Center} ed R Genzel and A Harris (Boston: Kluwer) 185 

\item Morris M and Serabyn E 1996 {\it Annual Review of Astronomy and Astrophysics} \textbf{34} 645

\item Nord M~E, Lazio  T~J~W, Kassim N~E, Hyman S~D, LaRosa T~N, Brogan C~L and 
Duric N  2004 {\it Astronomical Journal} \textbf{128} 1646 

\item Shibata K and Uchida Y 1987 {\it Proceedings of the Astronomical Society of Japan} \textbf{39} 559

\item Sjouwerman L~O, Kong A~K~H, Garcia M~R, Dickel J~R, Williams B~F, Johnson 
K~E, Primini F~A and Goss W~M 2005 {\it X-Ray and Radio Connections} ed L~O Sjouwerman and K Dyer, Published electronically by NRAO: \texttt{http://www.aoc.nrao.edu/events/xraydio}

\item Wang Q D, Gotthelf E V and Lang C C 2002 {\it Nature} \textbf{415} 148

\item Yusef-Zadeh F, Morris M, and Chance D 1984 {\it Nature} \textbf{310} 557

\item Yusef-Zadeh F and Morris M 1987 {\it Astrophysical Journal} \textbf{322} 721

\item Yusef-Zadeh F, Wardle M and Parastaran P 1997 {\it Astrophysical Journal} \textbf{475} L119

\item Yusef-Zadeh F, Hewitt J and Cotton W 2004 {\it Astrophysical Journal Supplement Series} \textbf{155} 421 

\end{thereferences}
\smallskip

\end{document}